Comprehensive Study on Influence of Low Dose Chronic Consumption of Silver Nanoparticles on Cognitive Functions and Identification of the Effect's Reasons

Anna A. Antsiferova * [1,2], Marina Yu. Kopaeva [1], Vyacheslav N. Kochkin [1], Pavel K. Kashkarov [1,2,3], Mikhail V. Kovalchuk [1,3,4]

[1] National Research Center "Kurchatov Institute", 1, Akademika Kurchatova sq., Moscow, 123182 Russia

[2] Moscow Institute of Physics and Technologies, 9, Institutskii lane, Moscow region, Dolgoprudny city, 141700, Russia

[3] Lomonosov Moscow State University, 1, GSP-1, Leninskiye gory, Moscow, 119991, Russia

[4] Shubnikov Institute of Crystallography, Federal Scientific Research Centre 'Crystallography and Photonics' of Russian Academy of Sciences, Leninskii pr. 59, Moscow 119333, Russia

* antsiferova_aa@nrcki.ru

Abstract

The influence of the daily prolonged administration of silver nanoparticles on cognitive functions of model mammals was studied. The accumulation of silver in whole brain, hippocampus, cerebellum, cortex and residual part of mice brain was investigated by highly precise and representative Neutron Activation Analysis and the histological studies were conducted in the present research.

Some of the main results of the comprehensive study are the description of the biokinetics of silver in the hippocampus, cerebellum and cortex, the finding of the jump increase of silver in the hippocampus and cerebellum after the period of administration of 120 days. Accumulation of silver in the mice brain and hippocampus itself as well as the integrity violation of the CA2 subregion caused long-term contextual memory impairment at 180 days of silver nanoparticle exposure. However, destruction of CA2 had started 60 days before the visible changes in behavior appeared.

Complex of the three types of the approach let us find the cognitive impairment and understand the reasons of it.

Since the beginning of the 2000s silver nanoparticles (Ag-NPs) have been used in various applications from pharmaceutics [1-4] to textile [5-7] and from hygiene products [8], packing materials [9] to daily food supplements [10-14]. Undoubtedly, Ag possesses unique antiseptic properties that were known since antiquity [15]. Therefore, people in the past successfully used silver dishes as antiseptics [16].

Nowadays silver is not forgotten and still competes with the modern antibiotics [17, 18].

Anyone may ask why not still silver but Ag-NPs? It must be taken into the account that any particles of nanosize demonstrate nature-likeness and the following effects. During the evolution nature developed mechanisms to deal with natural nanosized structures like proteins and cellular organelles. For instance, a living cell can recognize proteins in the native state [19], which are always nanosized, as solid or liquid food and uptakes them. Therefore, it can regard engineered nanoparticles with or without a corona of proteins as food or construction materials for the cell. The ways of uptake, such as phagocytosis [20] and micropinocytosis [21], of any nanosized structures have been developed by nature. This results in the of high cellular penetrability of engineered nanoparticles, their ability to accumulate inside cells and bioactivity.

For instance, in some cases Ag-NPs can accumulate in different organs such as liver, kidney, blood and brain of mammals [22-25]. Several works showed that all tissues except the brain easily remove silver [23,24] and some scientific researches reliably demonstrate the penetration of Ag-NPs via the blood-brain barrier and their jamming in the nerve tissue [23, 24, 26]. Therefore, brain can be regarded as a target organ for Ag-NPs.

There is a lot of scientific research that show toxicity of Ag-NPs to nerve cells as, at firsts, neurons, astrocytes and glial cells [27-35]. Of course, the toxic effect is almost straightly proportional to the doses of Ag-NPs and the exposure time.

Due to the accumulation of Ag-NPs in the brain and their toxicity to brain cells some the of recent works are devoted to study the influence of Ag-NPs on cognitive and behavioral functions of model mammals such as rats and mice [36-41]. Usually such researches are conducted for no more than 30 days and the doses of Ag-NPs are able to poison animals without acting on the brain [42].

The objective of the present research was the comprehensive studies of the influence of low concentrations of polyvinylpyrrolidone-coated (PVP-coated) nanoparticles on mammalian cognitive functions, quantitative description of the biokinetics of Ag-NPs in the brain, determination of the amounts of silver in the parts of the brain such as cerebellum, hippocampus, cortex and morphological studies of brain slices. All the studies have been conducted at the condition of prolonged 1 to 6 months oral administration of Ag-NPs, implemented with the use of modern high-quality equipment and priciest Neutron Activation Analysis (NAA).

Nanoparticles

Dynamical Light Scattering and Transmission Electron Microscopy demonstrated that the size of the nanoparticles was 34±5 nm and they had quasi-spherical shape. As it was shown [10] such PVP-coated nanoparticles are very stable (no significant size changes during one year of refrigeration) and can pass through the blood-brain barrier [22], which is due to the high stability of the Ag-NPs, their size and, perhaps, to their hydrophobic coating. Such hydrophobic coating provides an attachment to the cellular membranes and better penetrability to some types of the cells.

Influence of the Nanoparticles on Cognitive Functions

The influence of the Ag-NPs on behavioral and cognitive functions was described in our previous research [10]. It was shown that mammals go through three different periods such as anxiety, switching the mechanism of adaptation on and disturbance of cognitive functions. Perhaps, initial slight disturbance in behavior, which was thoroughly described in [10], is associated with the accumulation of silver in brains and the oxidative stress like one of the possible mechanisms of Ag-NPs cell influence. However, these initial changes do not seem to be fatal during the regarded period because of self-adaptation. However, the adaptation mechanism does not win in competition with the negative effect of Ag-NPs at the longer distance as it can be seen further.

Regarding long-term contextual memory, no changes in learning has been detected for any period.

Figure 1 demonstrates the effect of 30, 60, 120 and 180 day exposure of Ag-NPs on the contextual memory, in particular, the number of freezing acts during memory testing. The contexts were the electric pulses and the media of the chamber together for Ag-NPs groups of mice or just the media of the chamber for the control groups. It shows that there is no reliable difference between control groups and Ag-NPs groups for 30, 60 and 120 days. However, there is a tendency for 120 days that Ag-NPs mice made more freezing acts than the control mice. It seems that the adaptation mechanism is on at this period of Ag-NP administration because the memory seems even better. Also, it may indicate that the mice are somehow 'acute' during this time.

A reliable difference can be seen at 180 days of administration, when Ag-NPs mice made statistically a smaller number of the freezing acts than the control mice. It says about the degradation of long-term contextual memory which was due to the media of the chamber and secured by the very strong condition as electric pulses on mice feet. The mice just did not remmeber the negative feelings they were exposed to a day before testing.

Accumulation of Silver Nanoparticles in Brains and Their Parts

The amounts of silver accumulated in brains and their parts were studied by not so wide-spread technique for this purpose as NAA. It required several years to develop the approach of the best tracking, detecting and precise measuring of the amounts of each kind of nanoparticles, for instance, silver, in biological tissues [22, 43,44]. However, the principle of NAA is fairly simple. In NAA non-irradiated nanoparticles or other chemical compounds are administrated into living model organisms. In the required time their metabolism is stopped and the organisms are disassembled into organs. The organs containing the key element together with reference samples containing a known amount of the key element are irradiated in a nuclear reactor until the required activity of the key isotope has been achieved. After that the activity may be measured using gamma-spectrometry and calculated into the mass of the key element by comparing the activity with the activity of the reference sample.

NAA is the most sufficient, precise and sensitive (lower threshold is $10^{-9}$ g) method for measuring the amounts of nanoparticles in biological tissues. By the way, the technique is very representative (objects can be 5×10 cm, limited only by the geometry of channels) and do not require complicated sample preparation.

NAA has shown that the amount of silver in the control samples were below the critical recognition threshold so they were taken as zero. However, the amounts of silver in the brain and their parts of the experimental animals were determined with an error of no more than 10 %.

The accumulation of silver in the mice brains on the period of administration seems to be a monotonously increasing function striving to a constant value (Figure 2). The function proofs that silver do gradually accumulate in mice brain at the daily constant administration of Ag-NPs. This result is up to the similar data of our previous research [24].

Figure 3 shows the histograms of the silver concentration in the parts of the brain as a function of the period of administration. It can be observed that concentrations increase since the determined time in a jump-like way. The jump increases for the hippocampus (a) and cerebellum (b) can be seen from 120 days and from 180 days for the cortex and brain residue.

Figure 2 does not indicate any jump increase because brain consists of all the above parts and both of the jumps should be taken into the account. This is the reason why the function is more gradual.

Morphological Studies

Total observations at necropsy indicated that all the brains of the Ag-NPs exposed mice exhibited the expected anatomic features (e.g., characteristics of color, consistency, shape, and size) compared to their appearance to the control animals.

Analyses of the data obtained from measuring the bodyweight and brain weight of mice revealed that Ag-NPs had no significant effect either on the weight of the body or on the brain (Table 1).

By Nissl staining, no apparent structural difference was found in the whole brain, in particular, hippocampus, between Ag-NPs exposed and control mice in the '30 days' and '60 days' groups. We examined the amygdalar region of the brain for detection of changes. No differences in neuronal density were found in any of the experimental groups.

However, visible changes in the CA2 region of the hippocampus of Ag-NPs exposed mice compared to control mice in the '120 days' and '180 days' groups were observed. The stratum pyramidale was irregular and rarefaction in appearance. There was dispersed arrangement of neurons in the CA2 region of hippocampus after the exposure to Ag-NPs (Fig. 4).

Discussion

Thus, application of the three different approaches let us find out how Ag-NPs influence cognitive functions of mammals and to discover the reasons for the phenomena such as the accumulation of silver in the whole brain, its parts and the disturbance of the CA2 region.

NAA is undoubtedly a highly promising tool for nanoparticle biokinetics observation, that gives an opportunity to precisely measure the amounts of chemicals accumulated in organic and inorganic medium. In comparison to electron microscopy and mass spectrometry [42] it is very representative method because it lets us work with whole organs and their parts, obviating the needs for microsection required in the wider-spread techniques. Also, this is one of the reasons of its sensitivity, because it is easy to obtain gamma- or beta- signals from the whole sample such as an organ. Actually, sensitivity depends on the correct choose of the irradiation time, which is necessary to achieve the proper activity and, of course, on the properties of the radioisotope such as half-life and cross-section. For instance, the half-life of $^{110m}$Ag radioisotope is about 250 days, which lets us collect the activity of it within this period and even longer if the activity is not very high.

It should be noted that the NAA data obtained in our research presenting the amounts of silver accumulated in brain parts are unique and it is very interesting that two-time and even three-time jump increases were detected. Regarding to the previously obtained results on the excretion of silver from brain, which demonstrated the elimination rate increase after 120 days of Ag-NPs administration and earlier predicted changes in cell integrity, completely correspond to the present data [24]. Accumulation of silver in the brain part and jump increases may be due to the loosing of the cell integrity and increasing of the excretion speed. However, excretion speed increase may be regarded as losing the silver from the brain tissue. However, in this case the in-stream is also should be higher, which is proven by the growth of the silver mass in the whole brain and their parts. It is very logical that if the gates are open both side streams circulate in and out. Thus, in the competition between in- and out- streams, the first wins. The loosening of the CA2 subregion found at 120 and 180 days proofs the prediction about opening the gates.

It should be noticed that the CA2 region is responsible for the informational memory and recognition giving the opportunity to distinguish a certain media, situation and circumstances and to make the binds between them. Recent findings have shown that CA2 is involved in novelty detection [45,46], social memory [47], and ripple generation [48]. The CA2 has widespread anatomical connectivity, unique signaling molecules, and intrinsic electrophysiological properties. CA2 transmission does play a role in spatial learning, perhaps through its influence on overall network excitability [49].

Damage of CA2 would lead to the network disturbance, difficulties with binding the media and the situations, degradation of memory and to other unpredictable effects. It was shown in the present research that the damage to CA2 is accompanied by the long-term contextual memory reliable violation. Also, it is well known that the hippocampus is responsible for neurogenesis, therefore its damage may have a negative effect even on steam cells and slow the neurogenesis down. The results were found in the very 'strong' test, which is based on forming the fear of the environment by causing the pain. Detection of memory changes in this test definitely show evidence of cognitive damage. However, it should be noted that changes to the CA2 subregion together with the jump-like increase of silver content in the hippocampus were detected two months before the memory impairment has occurred.

It is clear that accumulation of silver in the brain and hippocampus itself as well as loosening of CA2 subregion led finally to the disturbance of the memory.

Since that, it is obvious that the daily consumption of low doses of Ag-NPs may be dangerous from different points of view: because of the possible memory and behavioral changes, which are due to the accumulation of silver in brain and their parts and to integrity changes in CA2 subregion of hippocampus.

Methods

A colloidal solution of Ag-NPs was provided by the food supplement Argovit C manufactured in the Russian Federation, Novosibirsk and was used as nanosilver. It was packed in an opaque plastic 0.5 l bottle. The nanoparticles were coated with PVP, the initial concentration of silver in the solution was 10 µg/ml in the distilled water.

Dynamical Light Scattering (DLS) (Malvern Zetasizer Nano ZS) and Transmission Electron Microscopy (TEM) (Titan 80-300, Thermo Fisher Scientific) were used to measure the size and stability of the nanoparticles. TEM has also been applied to visualize the nanoparticles. For this purpose, initial solution was dissolved by distilled water up to the concentration of 0.1 µg/ml in the amount of 15 ml. Half of the solution was used for the immediate measurements, the residual part had been preserved in the refrigerator at the temperature +2 ºC for 1 year and then investigated with DLS and TEM to check the stability of Ag-NPs.

For the immediate measurements the solution was divided into 2 equal parts for DLS and for TEM. Before the measurements the solutions were sonicated in an ultrasonic bath for 15 minutes. For DLS 1 ml of the solution per measurement was poured into a plastic cuvette and a series of measurements was made. For TEM 0.01 ml of the solution was applied to a carbon grid, dried and measured microscopically. The same experiment was accomplished with the preserved solution.

Eighty C57Bl/6 eight-week-old male mice, which weighted 19-27 g, had been purchased from the Stolbovaya supplier, Moscow, Russia. Mice were kept in the individual ventilated cages with the access to standard laboratory food and water ad libitum under with a room temperature controlled at 23 ± 2°C and 12/12 h. light/dark cycle. All experimental procedures were performed in the accordance with the rules of the Ministry of Health of the Russian Federation (№ 267 of 19.06.2013) and the Local Ethics Committee for Biomedical Research of the National Research Center 'Kurchatov Institute'.

Animals were randomly divided into four experimental groups by 20 mice in each group: '30 days', '60 days', '120 days' and '180 days'. Daily 10 animals in each group received orally Ag-NPs suspended in distilled water (50 μg per day) during the whole experiment. The rest of the animals received sterile water.

The body weight was monitored weekly during the whole exposure period.

To examine of the effect of silver nanoparticles on memory formation and retention mice were trained and their memory was tested in a contextual fear conditioning task using a Video Fear Conditioning System (MED Associates Inc.) and the computer program Video Freeze v2.5.5.0 (MED Associates Inc.). Video recordings of the animal behavior were made during training and testing. The number and duration of freezing acts were determined automatically.

Animals of each experimental group (the Ag-NPs exposed mice and control mice separately) were randomly assigned to two subgroups: 'fear conditioning (FC)' (n=6) and 'active control (AC)' (n=4). Mice from 'FC' subgroup were placed for 6 min into the experimental chamber, where they freely explored a new environment for 3 min, and then 3 electric foot shocks (1 mA, 2 s) were delivered with a 1 min interval followed by a 1 min rest period. Mice from the 'AC' subgroup were placed for 6 min into the experimental chamber, where they freely explored a new environment without a foot shock. Mice were returned to their home cages immediately after the training. 24 h after training animals were tested for long-term memory retention (3 min in the experimental chamber without a foot shock). Recollection of the chamber environment must have made the 'FC' group connect the chamber with the negative feeling of foot shocks, while 'AC' group could not make the connection with any negative action. The proportion of freezing acts versus test duration was assessed as a measure of long-term memory. Before placing each animal into the chamber, the chamber was wiped with 70% ethanol.

Animals were compared by the following parameters: the percentage of freezing acts in the range before and after the current was supplied during training, the percentage of freezing acts during testing.

Statistical analysis in the cognitive test was performed with GraphPad Prizm 6 by the nonparametric Mann-Whitney test. The differences were considered significant at $p < 0.05$. All data were expressed as means ± SEM.

After contextual memory testing of each experimental group animals were divided into 2 groups: mice for biokinetical studies (BS) (8 experimental and 8 control mice in each subgroup) and mice for histopathological studies (HS) (2 experimental and 2 control mice in each subgroup). Mice for biokinetics were divided into the2 following groups: 'whole brain biokinetics' (WBB) (4 experimental mice and 4 control mice) and 'brain parts biokinetics' (BPB) (4 experimental mice and 4 control mice). HS mice were anesthetized and perfused with 4% paraformaldehyde in phosphate buffered saline (pH 7.4). Brains were removed and postfixed. Coronal 40 μm sections were cut through the whole brain on Leica VT1200S vibratome. For HS every sixth section was Nissl stained with cresyl violet and examined by light microscopy (Zeiss Imager z2 VivaTome, Carl Zeiss, Germany).

All other mice were decapitated folowwing 0.6 % urethane anesthesia. BPB brains were postfixed with 1% paraformaldehyde during 24 hours, then weighted and divided on Leica VT1200S vibratome into hippocampus, cerebellum, cortex and brain residue. Then WBB and BPB samples were put into marked individual opened plastic boxes and dried at 75º C during 48 hours to the state of dried food.

Those samples in marked closed plastic boxes (Eppendorf) were put layer by layer in aluminum cases with the reference samples, where the amount of silver is known precisely, as shown in

Figure 5. Reference samples were prepared just prior to use. For this purpose state standard sample of silver with known mass amount of silver was poured onto sterile cotton wool and then reference samples with the same form-factor as to the experimental samples were prepared and put into plastic boxes. Also, a known amounts of state standard sample were poured onto paper rings, then dried and placed between the layers. Thus, one aluminum case contained one reference sample with the same form-factor as the experimental samples and 84 'flat' reference samples. The cases were put into vertical channels of the nuclear research reactor IR 8 (power is 8 MeW) and irradiated there for 24 hours at a neutron flux of $10^{-12}$ neutron/sec*cm$^2$ to obtain thenecessary activity of the $^{110m}$Ag radioisotope during NAA. After that all the WBB and BPB experimental and reference samples were examined by gamma-spectrometric elemental analysis (CANBERRA) to measure the activities of silver in the samples. The amount of silver in the experimental samples was calculated by comparison of the activities with the activities of reference samples, where the mass of silver was known precisely. The method error was neglected because of the low value. All data were expressed as means ± SEM.

Acknowledgments:

The present research was financially supported by the Russian Foundation for Basic Research (project No 16-32-00850).

Author contribution

A.A.A. came up with the idea of this research, she received the grant for it being the principal investigator, planned the scheme of the study, purchased all the necessary materials, performed the DLS and TEM studies with nanoparticles and wrote the paper, M.Yu.K.

Figure Legends

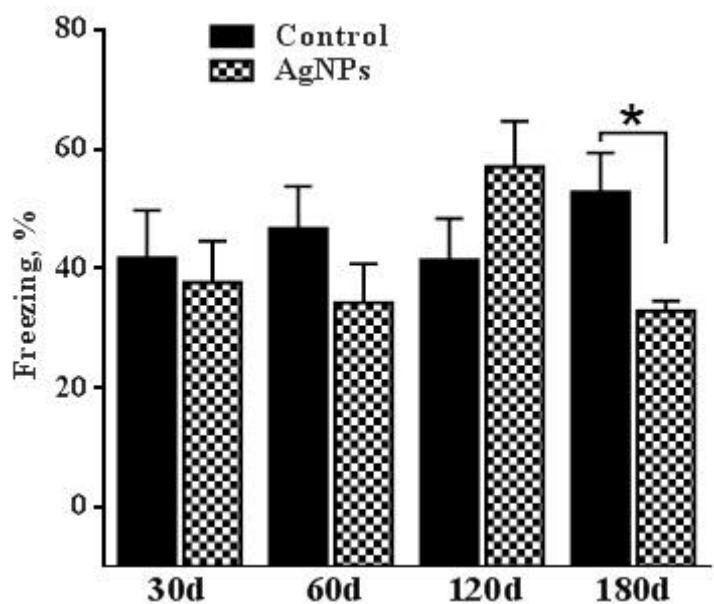

Fig. 1. Impairment of long-term contextual fear memory in the Ag-NPs exposed mice in the '180 days' group tested 24 h after training. The level of freezing was statistically lower in mice of the 'FC' Ag-NPs exposed group as compared with the 'FC' control group. This effect was not observed in the other groups. Data is represented as mean ± SEM. *p<0.05; nonparametric Mann-Whitney test (n = 6-8 animals per group).

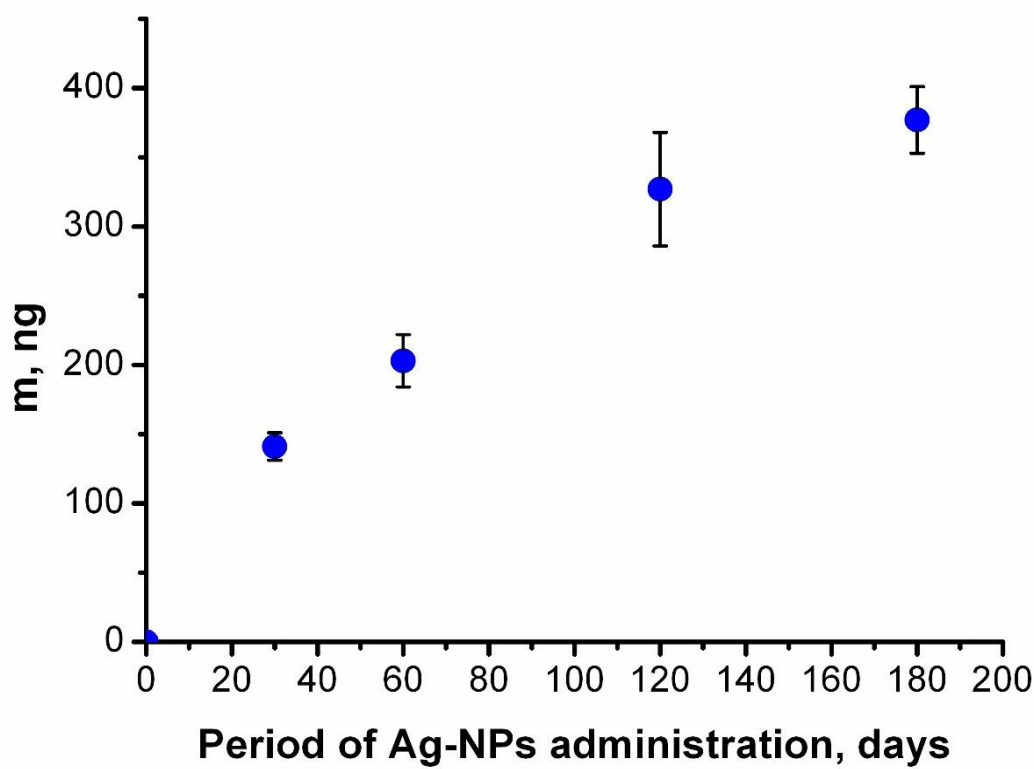

Fig. 2. Accumulation of silver in the brain. Dependence of the mass of silver in the brain on the period of administration of Ag-NPs obtained by NAA. It can be seen that silver is monotonically accumulated in brain with the time. Data is represented as mean ± SEM.

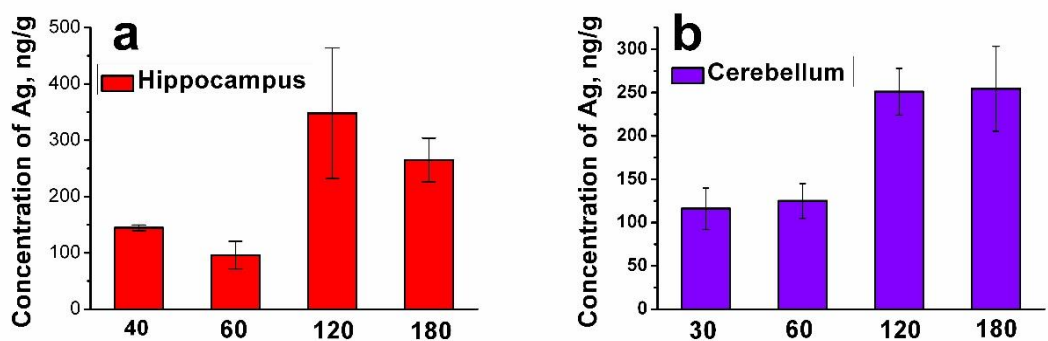

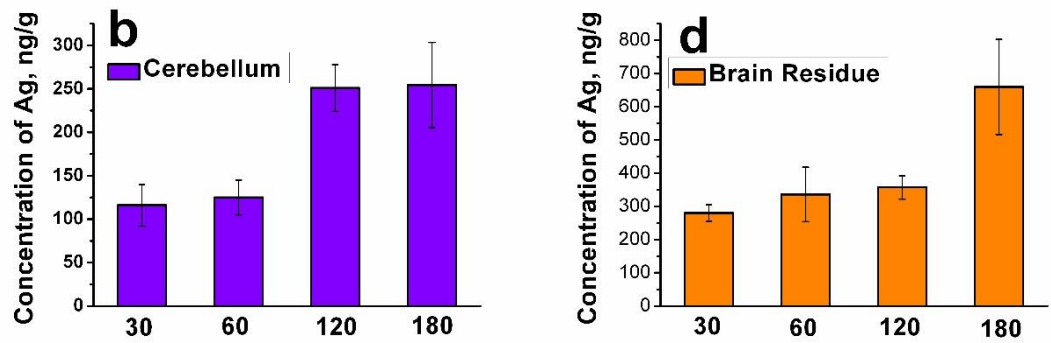

Figure 3. Accumulation of silver in the parts of brain: a) hippocampus, b) cerebellum, c) cortex, d) brain residue. Dependencies of the concentration of silver in the parts of the brain on the period of administration obtained by NAA in ng/g of the tissue. The jump increase in the silver concentration can be seen from 120 days of administration for the hippocampus and cerebellum and from 180 days for the cortex and brain residue. Data is represented as mean ± SEM.

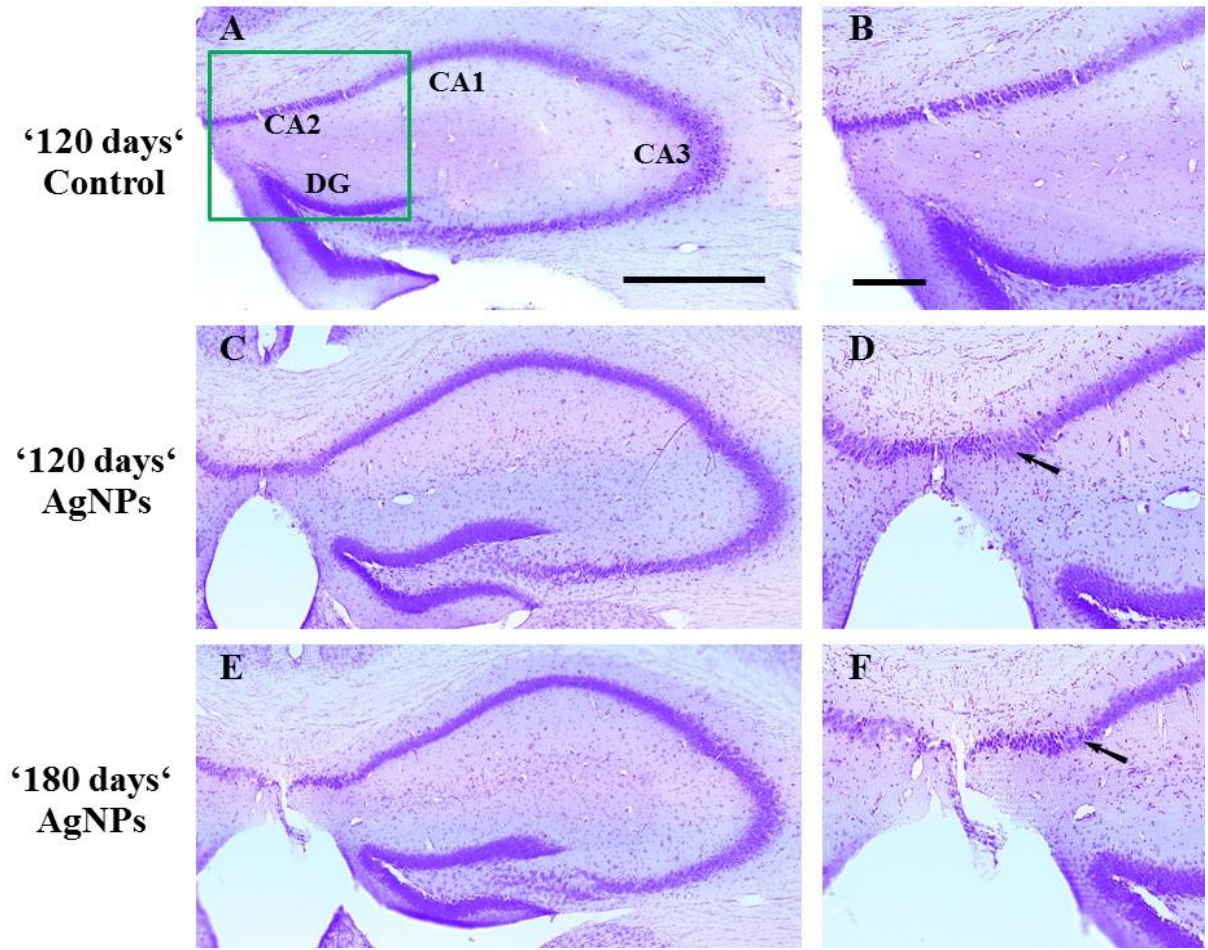

Fig. 4. Comparison of hippocampal structure of the mice. Nissl stained representative brain tissue sections: (A, B) control group, Ag-NPs exposed mice (C, D) '120 days' and (E, F) '180 days' groups. Visible changes in the CA2 region of hippocampus of Ag-NPs exposed mice compared to control mice (arrows). Scale bars: 500 µm (A, C, E), 200 µm (B, D, F).

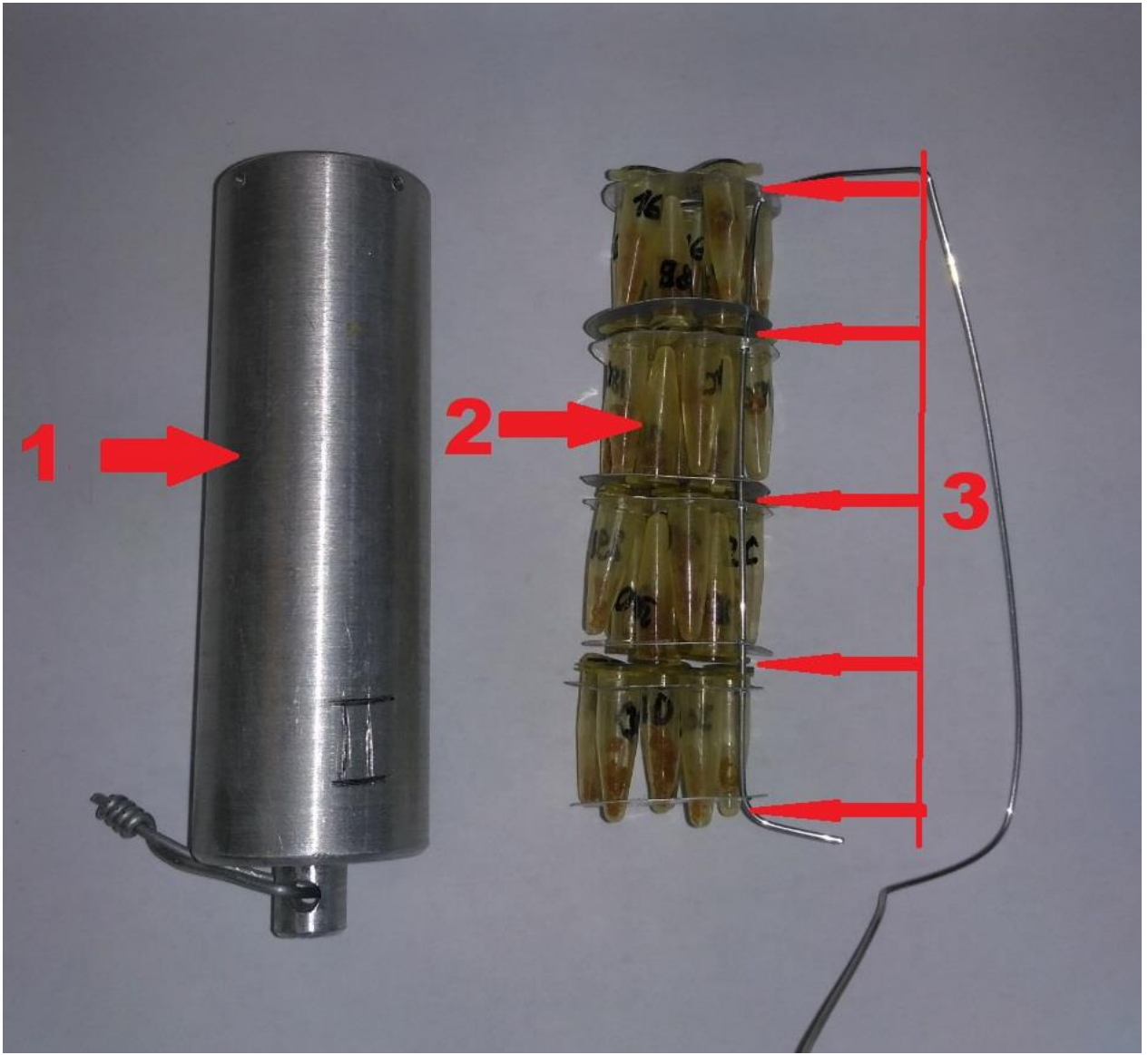

Figure 5. Apparatus for the NAA. Aluminum case (1) on the left, experimental samples in plastic boxes (2) and paper gaskets with reference samples (3) on the right.

Tables

Table 1. Effects of Ag-NPs exposure on body weight and brain weight of the mice. Data represent as mean ± SEM. No significant differences were detected. (n = 6 - 8 animals in each group).

| Groups | 30d | | 60d | | 120d | | 180d | |
|---|---|---|---|---|---|---|---|---|
| | Control | Ag-NPs | Control | Ag-NPs | Control | Ag-NPs | Control | Ag-NPs |
| Body weight (g) | 26,5±1,2 | 26,2±0,9 | 26,2±0,3 | 25,7±0,4 | 27,8±0,7 | 26,5±0,6 | 28,6±0,8 | 29,4±0,5 |
| Brain weight (mg) | 435±11 | 435±10 | 420±10 | 427±8 | 434±6 | 430±5 | 446±12 | 441±7 |